\documentclass{IEEEtran}

\usepackage{color}
\usepackage{acronym}
\usepackage{comment}
\usepackage{cite}
\usepackage{enumerate}
\ifCLASSINFOpdf
   \usepackage[pdftex]{graphicx}
\else
   \usepackage[dvips]{graphicx}
\fi
\usepackage[cmex10]{amsmath}
\usepackage{amsfonts}
\usepackage{mathrsfs}
\usepackage{array}
\usepackage{stfloats}
\usepackage[font=small]{caption}
\usepackage{subcaption}

\newcommand{\bs}[1]{\boldsymbol{#1}}
\newcommand{\wqss}{\bs{\omega}_{\rm \scriptscriptstyle QSS}}
\newcommand{\wu}{\bs{\omega_\upsilon}}
\newcommand{\wup}{\bs{\omega'_\upsilon}}

\newcommand{\voltage}{\bs{\upsilon}}
\newcommand{\vorticity}{\mathbf{w}}
\newcommand{\Clarke}{$\alpha\beta\gamma$}

\newcommand{\tout}{\mathscr{T} (\Gamma')}

\acrodef{rocof}[RoCoF]{Rate of Change of Frequency}
\acrodef{ufls}[UFLS]{Under Frequency Load Shedding}
\acrodef{ntg}[NTG]{National Transmission Grid}
\acrodef{ohl}[OHL]{Over Head Line}
\acrodef{tso}[TSO]{Transmission System Operator}
\acrodef{pmu}[PMU]{Phasor Measurement Unit}
\acrodef{qss}[QSS]{Quasi Steady State}
\acrodef{hv}[HV]{High Voltage}
\acrodef{ehv}[EHV]{Extra High Voltage}
\acrodef{sg}[SG]{Synchronous Generator}
\acrodef{hvdc}[HVDC]{High Voltage Direct Current}
\acrodef{vsc}[VSC]{Voltage Source Converter}
\acrodef{res}[RES]{Renewable Energy Source}
\acrodef{ac}[AC]{Alternate Current}
\acrodef{tn}[TN]{Transmission Network}
\acrodef{oem}[O\&M]{Operation \& Maintenance}
\acrodef{lom}[LoM]{Loss of Mains}
\acrodef{ibr}[IBR]{Inverter-Based Resource}
\acrodef{pll}[PLL]{Phase-Locked Loop}
\acrodef{ls}[LS]{Load Schedding}
\acrodef{qss}[QSS]{Quasi Steady-State}
\acrodef{coi}[CoI]{Center of Inertia}
\acrodef{dft}[DFT]{Discrete Fourier Transform}
\acrodef{fir}[FIR]{Finite Impulse Response}
\begin{document}
\title{Theoretical and Experimental Limitations of RoCoF Estimation}

\author{\IEEEauthorblockN{J.~Guti{\'e}rrez-Florensa\IEEEauthorrefmark{1},
F.~Sanniti\IEEEauthorrefmark{2},
D.~Tedeschi\IEEEauthorrefmark{3}, 
L.~Sigrist\IEEEauthorrefmark{4},
{\'A}.~Ortega\IEEEauthorrefmark{4} and F.~Milano\IEEEauthorrefmark{1}\\}
 \IEEEauthorblockA{\IEEEauthorrefmark{1} School of Electrical \& Electronic Engineering, University College Dublin,
Dublin, Ireland}
 \IEEEauthorblockA{\IEEEauthorrefmark{2} Department of Industrial Engineering,
 University of Padova, Italy}
 \IEEEauthorblockA{\IEEEauthorrefmark{3} Terna S.p.A.,
  Roma, Italy}
 \IEEEauthorblockA{\IEEEauthorrefmark{4}Institute for Research in Technology, ICAI,
 Comillas Pontifical University, Madrid, Spain}
 }

\maketitle

\begin{abstract}
A precise estimation of the \ac{rocof} is crucial for secure power system operation.  In fact, \ac{rocof} is strictly related to the amount of the available physical and/or virtual inertia of the system and the severity of the active power unbalance following a disturbance.  For this reason, it is widely exploited in different protection systems, \textit{e.g.}, Anti-Islanding, \ac{ufls} and wide-area protection systems.  
The new paradigm of modern power systems, with a low-inertia and converter-based generation assets, is increasing the transient severity, making the frequency and the \ac{rocof} estimation more complex and less precise for the actual devices.  
This work addresses this issue by proposing a numerically robust approach based on concepts inherited from differential geometry and fluid mechanics.  The proposed approach is then tested with high-sampling real experimental measurements and used to develop a faster control logic for a \ac{rocof}-based \ac{ufls} control scheme.  The proposed approach provides information to protections regarding the nature of the contingency which can be used to improve its response.
\end{abstract}


\section{Introduction}

\subsection{Motivation}

\ac{rocof} is formally defined as the time derivative of the instantaneous frequency, \textit{i.e.} the second time derivative of the phase angle of a voltage signal \cite{IEEE2018synchrophasors}.
In practical applications, it is computed starting from the estimation of the instantaneous frequency, typically provided by a \ac{pll} or from the fundamental frequency estimated by a \ac{pmu}.  Then, the time derivative is computed with the adoption of some filtering and its value is averaged on a rolling time window for a smoother behavior \cite{entso2018rocof}.  In addition, for protection schemes, an intentional time delay between the detection and actuation can be used to avoid undesirable trips.

Extended rolling windows are typically adopted to improve the \ac{rocof} estimation by reducing noise, unbalances and distortion, at the expense of the promptness of the relay.  On the other hand, a shorter time window improves the responsiveness, but oscillations in the \ac{rocof} estimation, for instance, can lead to undesirable trips.  

A recent interpretation of the frequency in the framework of the differential geometry (see \cite{milano2022geometrical}) creates the conditions to approach the actual \ac{rocof} computation issues from a novel perspective.  In particular, contributions in \cite{gutierrezflorensa2025qss} are especially relevant as they define \ac{qss} frequency - a novel quantity that captures the slowly-varying fundamental value of frequency - and the time derivative of circulation - a metric that evaluates the physical existence of such frequency.  Together, these concepts are here exploited to assess the theoretical and experimental limitations of \ac{rocof} and introduce a novel approach for its estimation.

\subsection{Literature Review}

Nowadays, the problem of the \ac{rocof} estimation in power system applications is still an open challenge.  Low-inertia systems exhibit higher values of \ac{rocof} during transients and, at the same time, its estimation is harder due to the higher distortion of the electrical quantities \cite{rocof_meas}.

At present, there is no standard approach to compute \ac{rocof} in power systems.  Several techniques have been proposed in the last years \cite{rocof_review}, which can be grouped in two main categories depending on how the frequency is computed: (i) based on instantaneous frequency measurement  (provided, \textit{e.g.}, by a \ac{pll}  or a \ac{fir} filter \cite{rocof_pll}), and (ii) based on fundamental frequency measurement (exploiting, \textit{e.g.}, recursive \ac{dft} techniques \cite{phasor_fdt, dft'19}).  As the \ac{pll}-based estimation is sensitive to amplitude steps, noise and harmonics, typically the formal estimation of \ac{rocof} is filtered and then averaged on a rolling time window, introducing latency in the estimation \cite{rocof_step}.

\ac{dft}-based techniques can be faster since they elaborate on a phasor measurement, even if they require by definition a given time window for the estimation \cite{dft'19}.  These methods are also less sensitive to noise and harmonics but their precision deteriorates especially in the first instant after a disturbance, where the signal distortion is higher \cite{rocof_syncrophasor_ULFS}.
Hence, also in this case a proper averaging window has to be selected, vanishing the promptness of the algorithm \cite{testrocof}.

In recent literature, \ac{pmu}-based techniques for \ac{rocof} estimation have arisen among the others, because of their fast reporting rates and responsiveness, and the wide monitoring ability of the \ac{pmu} architecture \cite{pmu_rocof}.  \ac{pmu}-based \ac{rocof} estimation techniques do not employ averaging to limit the uncertainty given by noise, distortion, and unbalances.  As a consequence, the IEEE has significantly relaxed the \ac{rocof} error requirement from 0.01 Hz/s to 0.4 Hz/s \cite{rocof_std}.  In practice, this level of uncertainty make the \ac{rocof} estimation useless for typical applications, since values close to 1 Hz/s can already be considered critical for the stability of the system \cite{entso2016criteria}.  Several synchrophasor-based algorithms have been compared in \cite{pmu_rocof} and \cite{testrocof}, where it becomes evident that \acp{pmu} provide optimal performance in presence of low fluctuations, as typical of inter-area oscillations, but their measurement is hampered during fast transient events.  

The deployment of \ac{rocof}-based relays is mainly focused on three applications: (i) island detection for distribution networks \cite{rocofislading}, (ii) protection from \ac{lom} of embedded generators \cite{rocofLoM} and (iii) \ac{rocof}-based \ac{ufls} schemes, as part of the power system defense plan of many \acp{tso} \cite{entso2010technical}.  \ac{rocof}-based \ac{ufls} schemes allow improving their responsiveness by anticipating the decision to shed load \cite{UFLS_COI}, \cite{kundur2007power}.  
The latter application has gained attention in recent years due to the increased penetration of \acp{ibr}, which is worsening the effects of power disturbances on frequency stability
, as recent blackout events have proven \cite{spain,chile}.

Among several possible topologies of \ac{ufls} schemes \cite{ulfs_scheme}, traditionally based on frequency measurements, semi-adaptive \ac{ufls} schemes introduce \ac{rocof} thresholds to trigger the relays in addition to frequency thresholds.  For \ac{ufls} application, larger absolute \acp{rocof}  values can lead to the following major issues sorted by criticality:
\begin{itemize}
\item Frequency collapse before the relays have time to trip;
\item Unintentional tripping due to inaccuracy in the measurement of the frequency and/or \ac{rocof};
\item Over extensive \ac{ls} (due to the time delay of activation of the relays).
\end{itemize}

A trade-off should be found between the promptness of the intervention and the prevention of unintentional trips.  In practical applications, often large time windows are chosen, ranging from 500 to 1000 ms \cite{entso2020rocof}.  ENTSO-E, \textit{e.g.}, recommends 500 ms as a rolling time window for adequate \ac{rocof} monitoring \cite{entso2018rocof} and recommends also a maximum  disconnection time delay \cite{entso2010technical}.  
However, to date, recommended criteria for a suitable definition of threshold values and tripping logic are still missing for \ac{rocof}-based \ac{ufls}.

\subsection{Contributions}
Based on a recent development proposed in \cite{gutierrezflorensa2025qss}, the paper further elaborates on the definition of \ac{rocof} by means of \ac{qss} frequency and utilizes a metric, namely the time derivative of the circulation, able to determine what parts of the voltage trajectory following a power system transient retain the meaning of a slowly-varying fundamental frequency.  

Specific contributions of this work are as follows.
\begin{itemize}
\item Derive an expression for the \ac{rocof} based on the derivation of \ac{qss} frequency;
\item Validate the proposed expression by means of high-sampling real-world data provided by the Italian TSO; 
\item Discuss the impact of the proposed \ac{rocof} estimation on the tripping logic of a semi-adaptive \ac{ufls} scheme.
\end{itemize}

\subsection{Paper Organization}
This paper is structured as follows.  Section \ref{sec: background} recalls relevant concepts for the development of the paper and introduces a novel approach for \ac{rocof} estimation.  Section \ref{sec:real_data} assess a real-world case scenario to validate the findings in the previous section.  Section ~\ref{sec: app} evaluates the potential application of the presented theory in \ac{ufls} protections through an specific case study.  Section \ref{sec:remarks} provides final remarks and discusses practical considerations.  Finally, Section \ref{sec: conclusions} summarizes the main findings and conclusions of the paper.

\section{Background and \ac{qss}-Based \ac{rocof} Definition}\label{sec: background}

Based on recent works (see \cite{milano2022geometrical, milano2025lagrange, gutierrezflorensa2025qss}), we adopt the differential geometry framework to define electrical quantities in direct analogy with fluid mechanics.  On this basis, this section recalls the relevant definitions of \textit{\ac{qss} frequency} and \textit{time derivative of circulation} from \cite{gutierrezflorensa2025qss}, for the developments of the paper, and provides a novel definition of \ac{rocof} based on this theoretical background.

In the context of differential geometry, paper \cite{milano2022geometrical} defines geometric frequency multivector, $\hat{\Omega}_{\voltage}$, as an invariant composed by a translation represented by a scalar, $\rho_{\voltage}$, and an axial vector, $\wu$, that represents a rotation:
\begin{equation}
    \hat{\Omega}_{\voltage}=\rho_{\voltage}+\wu
    \,,
\end{equation}
for:
\begin{equation}
\begin{array}{ccc}
    \rho_{\voltage}=\dfrac{|\voltage\cdot\voltage'|}{|\voltage|^2} & ; &\wu=\dfrac{|\voltage\times\voltage'|}{|\voltage|^2}=\kappa|\voltage|
    \end{array}\,.
\end{equation}
Where the latter term closely matches the instantaneous frequency obtained with a conventional \ac{pll} as concluded in \cite{milano2022paradoxes}.

In \cite{milano2025lagrange}, an analogy between geometric frequency and Lagrange's derivative is used to decompose $\wu$  in different terms with precise physical meanings.  These meanings come from the \textit{fundamental theorem of the kinematics of continua} that states that an arbitrary instantaneous state of motion is at each point a uniform velocity of translation, a motion of extension, a shearing motion and a rigid rotation \cite{truesdell1954kinematics}.  Among these, for the purposes of this work, the component of interest is the rigid rotation quantified by half the vorticity, $\vorticity$, \textit{i.e.} the curl of voltage ($\nabla\times\voltage$), as the one that captures the meaning of fundamental frequency.

However, the estimation of vorticity, in this terms, requires to know the relationship between voltages and fluxes which is not possible in practice.  For this reason, the work in \cite{gutierrezflorensa2025qss} uses a Cauchy's theorem to define the \ac{qss} frequency, $\wqss$, as an equivalent quantity of vorticity in fluid mechanics which, for any given period $T$:
\begin{equation}
    \wqss=\frac{1}{T}\oint_T \wu \,d\tau =\frac{1}{2}\vorticity \, ,
\end{equation}
for any given period $T$ where the second identity is the result of Cauchy's theorem, when no torsion is assumed.
The quantity $\wqss$ thus conceptually bridges instantaneous and Fourier's frequency by retaining the slowly-varying fundamental frequency meaning while being valid not only during ideal balanced stationary conditions.

The election of period $T$ is not trivial and is determined by the aforementioned Cauchy's theorem applicability condition.  Cauchy's theorem is valid for regular curves that fulfill:
\begin{equation}
\kappa_T=\oint_{s_T}\kappa\,ds=\oint_T\kappa|\voltage|\,d\tau=\oint_T|\wu|\,d\tau=2\pi\,,
\end{equation}
where $\kappa_T$ and $s_T$ are the total curvature and arc length respectively.  Note that this is, in fact, the necessary condition, for a space curve, to describe a Jordan's curve, \textit{i.e.} a simple closed curve \cite{Stoker1969differential}.
Thus, we define the period as:
\begin{equation}
  T=\text{inf}\,\left\{t: t>t_0\, , \int_{t_0}^{t+t_0}|\wu| \, d\tau= 2\pi \right\} \,.
  \label{eq:Tdef}
\end{equation}

Period, by means of \eqref{eq:Tdef}, is defined through the voltage trajectory as the time it takes, for a given initial time $t_0$, to close the described curve.  Alternatively, if the trajectory does not draw a closed curve, period and, consequently, frequency cannot be defined.

In summary, estimation of \ac{qss} frequency requires to assume that the trajectory describes a Jordan curve.  For this reason, \cite{gutierrezflorensa2025qss} defines a metric, namely time derivative of circulation, $\Gamma'$, independent from condition \eqref{eq:Tdef}, to determine wether this assumption is satisfied.  Voltage trajectory describes a closed curve and, hence, \ac{qss} frequency retains the fundamental frequency value of the signal if:
\begin{equation}
  \Gamma' = \oint_{T}(|\voltage|^2)'\, d\tau = 0 \, ,
  \label{eq:Kelvin}
\end{equation}
whereas if $\Gamma'\neq0$, voltage trajectory is not closed and \ac{qss} frequency measurement lacks of physical meaning associated with periodicity.  This condition is derived from the analog definition of circulation in fluid mechanics ($\Gamma=\oint_T|\voltage|^2\,d\tau$) and the application of Kelvin's circulation theorem.

As in practice it is not realistic to expect to obtain exactly zero from \eqref{eq:Kelvin} due to measurement and numerical errors, we compare $\Gamma'$ to a threshold, $\epsilon$, that allows us to discriminate if the non-zero outcome comes from unavoidable practical errors or due to the loss of periodicity.  Hence, at any instant $t$, the voltage is assumed to be periodic and $\wqss$ to be defined if:
\begin{equation}
  |\Gamma'| \leq \epsilon \,.
  \label{metric}
\end{equation}

Then, \ac{rocof} can be obtained as the time derivative of \ac{qss} frequency subject to condition \eqref{metric}:
\begin{equation}
    \text{RoCoF}=\wqss'\, , \qquad \text{if} \; |\Gamma'| \leq \epsilon \, .
    \label{eq:wqss_prime}
\end{equation}
However, in real-world applications, \ac{rocof} calculation involving a derivative of $\wqss$ can be prone to noise and numerical issues.  For this reason, in practice, \ac{rocof} is normally estimated as an average value of a mobile window, within a time span, $\Delta t_w$, to filter out oscillations, distortions, etc.  \cite{entso2020rocof}.  However, such window might include “wrongly” measured frequency values such as spikes during the transient.  

In this work, we propose a novel \ac{rocof} estimation approach that derives the rate of change of \ac{qss} frequency and does not take into account the time intervals where frequency lacks physical meaning, namely, condition \eqref{metric} is not satisfied.
To do so, we first introduce a \textit{tout function}, $\tout$, that substantially implements a conditional boolean value:
\begin{equation}
    \tout=\frac{\text{sign}(\epsilon-|\Gamma'|)+1}{2}\,,
    \label{tout}
\end{equation}
that returns a $1$ if condition \eqref{metric} is satisfied and $0$ otherwise.

We can then define the average \ac{rocof} estimation as:
\begin{equation}
    \boxed{\overline{\text{RoCoF}}=\frac{1}{\Delta \tilde{t}_w}\int_{t-\Delta t_w}^{t}\wqss'\tout\,d\tau}
    \label{eq:rocof}
\end{equation}
where $\Delta \tilde{t}_w = \int_{t-\Delta t_w}^{t} \tout \, d\tau$.  The main feature of this \ac{rocof} estimation approach is that periods where frequency is not defined do not contribute to the calculation of the average.  Moreover, the tout function provides information  of the nature of any sudden frequency variation, which can be used to adapt the average time window for a faster estimation without loss of precision.

\color{black}

\section{Validation through Measurements}\label{sec:real_data}

\begin{figure}[htb]
  \centering
  \includegraphics[scale=0.85]{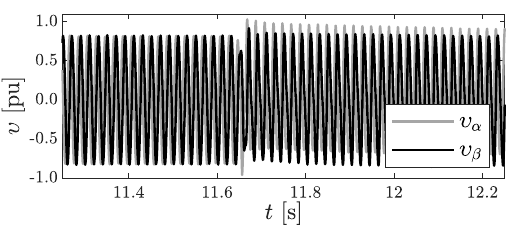}
  \caption{Measurements of the normalized voltage measurements during the transformer energisation event in the \Clarke{} frame.}
  \label{trafo_energ_voltages}
  \vspace{-2mm}
\end{figure}

In this section, expression \eqref{eq:rocof} is applied to high-sampling-rate measurements in order to compare it with the conventional \ac{rocof} computation from instantaneous frequency.  The available recordings, with a sample rate of $5$kHz, were provided by the Italian \ac{tso} (Terna S.p.A.) and are related to a transformer energisation during a power system restoration test performed on a region of the Italian Transmission Network\footnote{Note that frequency and \ac{rocof} relays have been disabled for the purpose of the test as the transformer energisation leads to significant \ac{rocof} deviations during the first instants due to the small size of the network.}.  The measured voltage time response related to the event is shown in Fig.~\ref{trafo_energ_voltages}.

\begin{figure}[htb]
  \centering
  \includegraphics[scale=0.825]{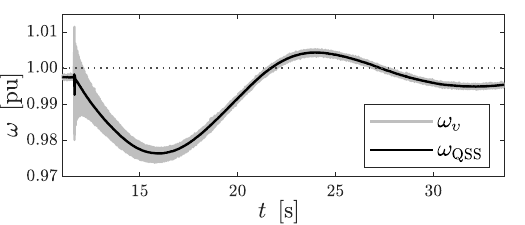}
  \caption{Instantaneous frequency, $\wu$, and \ac{qss} frequency, $\wqss$, results of the transformer energisation case study.}
  \label{trafo_energ_wv_W}
    \vspace{-3mm}
\end{figure}

Instantaneous frequency estimation, $\wu$, from \ac{pll} has been filtered through a first-order Butterworth filter.  Both instantaneous and \ac{qss} frequency results are shown in Fig.~\ref{trafo_energ_wv_W}.  Results show that the latter filters out the oscillations under stationary conditions while  damping the spikes in the transient.  This behavior is consistent with the findings in \cite{gutierrezflorensa2025qss}.

\begin{figure}[htb]
  \centering
  \includegraphics[scale=0.825]{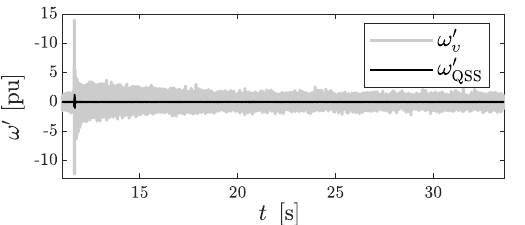}
  \caption{Time derivative of instantaneous and \ac{qss} frequency, $\wup$ and $\wqss'$ respectively, of the transformer energisation case study.  }
  \label{filtered_wv_w_prime}
    \vspace{-2mm}
\end{figure}

The immediate consequence of these results is that the formal estimation of the \ac{rocof} by means of the \ac{qss} frequency in \eqref{eq:wqss_prime} is less sensitive to noise than conventional instantaneous frequency estimation, as shown in Fig.~\ref{filtered_wv_w_prime}.

\begin{figure}[htb]
  \centering
  \subfloat[]{\includegraphics[scale=0.9]{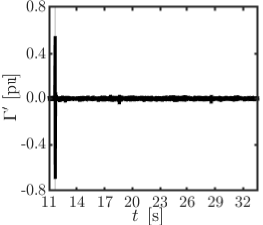}}\label{real wv}
  \subfloat[]{\includegraphics[scale=0.9]{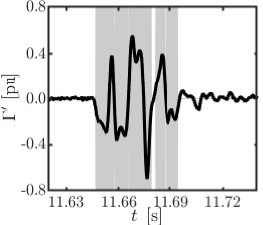}}\label{real circ}
  \caption{Overview of time derivative circulation, $\Gamma'$, results of the transformer energization case study (a), and detailed view of the results under the event (b).  Periods where condition \eqref{metric} is not satisfied are gray shadowed.}
  \label{trafo_energ_Gamma'}
    \vspace{-2mm}
\end{figure}

As previously stated, a rolling average can be applied to the formal estimation of the \ac{rocof} and therefore, the estimation in \eqref{eq:wqss_prime} is compared to the one proposed in \eqref{eq:rocof}, requiring satisfying the circulation condition.  With this in mind, we first assess the time derivative of circulation.  We set the threshold $\epsilon=0.05$ by considering the noise in the measurement under stationary conditions.  Results show that condition \eqref{metric} is not satisfied during the energization transient (see Fig.~\ref{trafo_energ_Gamma'}) and, hence, frequency (and thus \ac{rocof}) estimation lacks of physical meaning.

\begin{figure}[htb]
  \centering
  \includegraphics[scale=0.85]{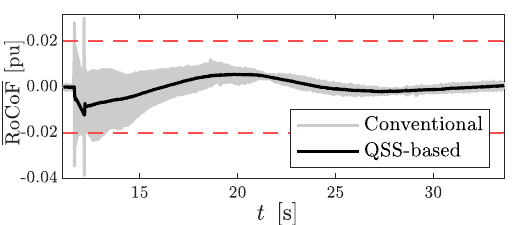}
  \caption{\ac{rocof} estimation as an average of time derivative of instantaneous frequency (Conventional) and by definition in \eqref{eq:rocof} (QSS-based) for $\Delta t_w=500$ms.  Red-dotted line represents a realistic threshold for system stability control.  }
  \label{fig:measrocof}
    \vspace{-2mm}
\end{figure}

By knowing the instants of satisfying the circulation condition, we finally estimate the \ac{rocof} as an average value of a rolling window of $\Delta t_w=500$ ms, as recommended by ENTSO-E (see \cite{entso2018rocof}), for both instantaneous frequency measurement and the proposed approach in \eqref{eq:rocof}.  Figure~\ref{fig:measrocof} depicts the results.  Red dashed lines represents \ac{rocof} values of $\pm1$ Hz/s, namely, the minimum values considered potentially critical for the stability of the system \cite{entso2016criteria}.  

In the first instants after the disturbance, the \ac{rocof} computed by the instantaneous frequency exceeds the critical values, whereas the \ac{qss}-based \ac{rocof} stays within them (as it should be in this case).  Moreover, the spikes during the transient might lead to the wrong conclusion that frequency is increasing, whereas the proposed approach is capable of smoothing that response by providing a more accurate estimation of the \ac{rocof} as the time derivative of frequency.

The inherently smoother response of the proposed approach allows us to shorten the size of the rolling time window, $\Delta t_w$, of the \ac{rocof} estimation.  Conventional estimation requires longer time windows to ensure noise and transient spikes are averaged out.  Estimation by means of \ac{qss} frequency, excluding periods in which frequency and, consequently, \ac{rocof} are not defined, permits a less restrictive methodology.

\begin{figure}[htb]
  \centering
  \includegraphics[scale=0.85]{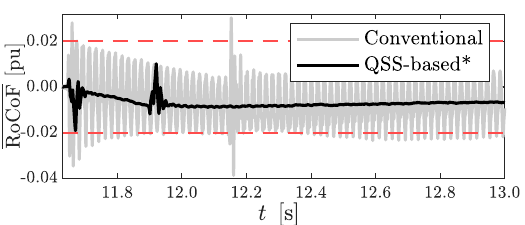}
  \caption{Detailed view of \ac{rocof} estimation as an average of time derivative of instantaneous frequency for  $\Delta t_w=500$ ms (Conventional) and by definition in \eqref{eq:rocof} with a reduced time window, $\Delta t_w=250$ ms (QSS-based*).  Red-dotted line represents a realistic threshold for system stability control.}
  \label{fig:measrocof_250ms}
    \vspace{-2mm}
\end{figure}

To support this conclusion, Fig.~\ref{fig:measrocof_250ms} plots the results of conventional \ac{rocof} estimation for $\Delta t_w=500$ ms and by means of \eqref{eq:rocof} with $50\%$ shorter rolling window.  Results prove that, although reducing the averaging interval, the \ac{qss} based \ac{rocof} estimation is less affected by noise and still remains within the normal operating range.

\section{Applying QSS-based RoCoF to UFLS}\label{sec: app}

In this Section, we present a potential application aimed to improve the \ac{rocof} estimation of \ac{rocof}-based protection relays in \ac{ufls} schemes.
For the purpose of this study, a semi-realistic scenario is set up starting from the IEEE 39-bus benchmark system, considering the dynamics of the lines, to simulate a large outage and the subsequent activation of a two-step \ac{rocof}-based \ac{ufls} scheme.  
The \ac{ufls} scheme has been implemented according to the simulation platform presented in \cite{rocof500}, which can represent a benchmark for testing the effectiveness of \ac{rocof} estimation in power system applications.  

The IEEE 39-bus system has been modified to emulate a low-inertia system.  
In particular, a set of \acp{ibr} are added at every generation bus.  The \ac{ibr} model employed and its primary controllers are described in \cite{DER}.  At generation buses, the power share of \acp{ibr} is imposed through the parameter $\gamma \in [0, 1]$, which scales the capacity of both synchronous machines and \acp{ibr} to ensure that the total generation capacity $S_{\rm n}$ remains constant:
\begin{equation}
    \begin{array}{ccc}
      S^{\rm IBR}_{\rm n} = S_{\rm n} \gamma &;&
  S^{\rm Syn}_{\rm n} = S_{\rm n} (1-\gamma)
    \end{array} \,,
\end{equation}
where $S^{\rm IBR}$ and $S^{\rm Syn}$ refer to the apparent power delivered by \ac{ibr} and synchronous resources respectively.

Inertia constants, $H$, power rates, ${S_n}$, and synchronous and transient reactances, $x_{\rm d}$, $x_{\rm d'}$, $x_{\rm q}$, $x_{\rm q'}$ of each synchronous machine are scaled accordingly.  \ac{ibr} models and controllers utilized in this case study include current limiters and anti wind-up limiters.  With these aforementioned assumptions, the modified 39-bus system is configured as a low-inertia system.

\begin{figure}[htb]
  \begin{center}
    \includegraphics[width=0.65\linewidth]{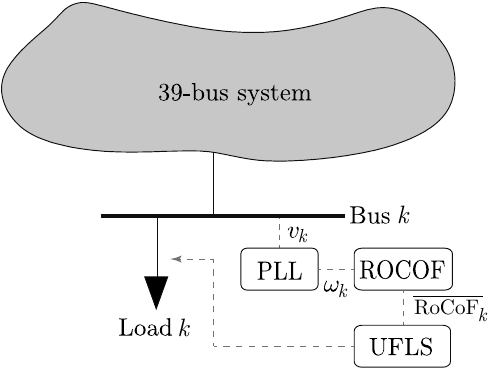}
    \caption{Diagram of the \ac{ufls} scheme for each load bus $k$ of the IEEE 39-bus modified system.}
    \label{fig:39ufls}
  \end{center}
  \vspace{-3mm}
\end{figure}

A two-step semi-adaptive \ac{ufls} scheme has been applied to all buses where a load is connected.  Figure~\ref{fig:39ufls} shows the diagram of the \ac{ufls} scheme acting on the $k^{th}$-load.  We assume here that a \ac{pll} computes the instantaneous frequency, $\omega_k$, starting from the bus voltage $v_k$, which is then passed through a low-pass filter and a washout filter by the \ac{rocof} block to compute $\overline{\rm RoCoF}_k$.  The way in which $\overline{\rm RoCoF}_k$ is computed can be different if we consider the enhancement given by the derivation of $\Gamma'$ ---see \eqref{eq:rocof}.

\begin{figure}[htb]
    \centering
 \includegraphics[width=0.9\linewidth]{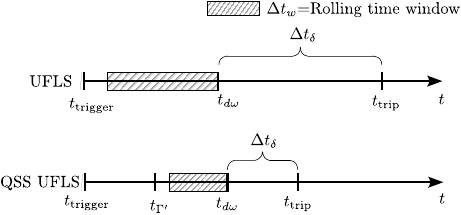}
    \caption{Tripping logic of the implemented \ac{ufls} schemes.}
    \label{fig:shedlogic}
    \vspace{-2mm}
\end{figure}

Figure~\ref{fig:shedlogic} details the two tripping logic schemes implemented in the \ac{ufls} block adopted for the simulations.  $t_{\rm trigger}$ refers to the time when a transient event occurs, $t_{d\omega}$ to the time when the rolling window detects a \ac{rocof} value exceeding a threshold and $t_{\rm trip}$ to the time when the protection trips.  

We consider a conventional tripping logic, labeled as \textbf{\ac{ufls}}, characterized by a large rolling window, $\Delta t_w$, and a large time delay, $\Delta t_{\delta}$, according to the recommendations given in \cite{entso2018rocof,entso2010technical} and the simulation performed in \cite{rocof500}.  Then, we propose a tripping logic based on the considerations expounded in the previous sections, named \textbf{QSS \ac{ufls}}, where we are able to exclude from the rolling time window $\Delta t_w$ the first instants after a trigger event where the frequency has no physical meaning associated with periodicity, until time $t_{\Gamma'}$.  Since we are sure that, starting from that instant, what we are measuring is a signal with a defined period, we can afford to shorten  $\Delta t_w$.  In this way, the relay issuing is faster but without the risk of unintentional tripping.

\begin{figure}[htb]
    \centering
    \includegraphics[width=0.7\linewidth]{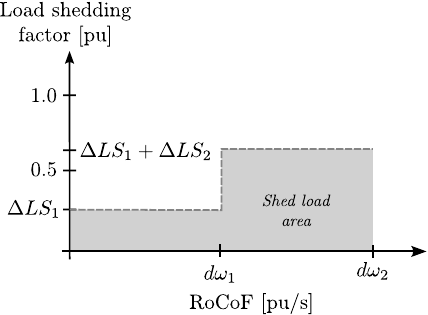}
    \caption{Tripping thresholds of the implemented \ac{ufls} schemes.}
    \label{fig:shed}
    \vspace{-2mm}
\end{figure}

Regarding the \ac{ufls} tripping scheme, it consists of two \ac{rocof} thresholds, $d\omega_1$ and $d\omega_2$ which command the disconnection of  two \ac{ls} portions, namely $\Delta LS_1$ and $\Delta LS_2$ respectively.  Fig.~\ref{fig:shed} shows the tripping scheme adopted for the simulations.  In this specific scenario, the purely frequency thresholds are not implemented on purpose, since the aim is to analyze the effect of \ac{rocof} measurements on the \ac{ufls} scheme and so on the dynamic performance of the system.  

A large generation outage, involving both synchronous generators and \ac{ibr} installed on buses 1, 2, 8, 10 is simulated, for a total of $2.44 \, \text{GW}$ of power outage.  The sum of the two thresholds of all \ac{ufls} disconnects a total of $2.43 \, \text{GW}$ of load, balancing the initial outage.  Parameter values utilised in the simulations are shown in Table \ref{tab:data}.

\begin{table}[tbh!]
  \renewcommand{\arraystretch}{1.3}
  \centering
  \caption{Data setting for the \ac{ufls} scheme of Fig.  ~\ref{fig:39ufls}.}
  \resizebox{\columnwidth}{!}{%
  \begin{tabular}{c|l|c|c}
    \hline
     \bf{Param.} & \bf{Description} & \bf{UFLS} & \bf{QSS} \\
                 &                  &   & \bf{UFLS} \\
     \hline
    $K_i$ & [-] PLL integral gain                                          & 0.03 & 0.03  \\
    $K_p$ & [-] PLL proportional gain                                      & 0.2  & 0.2   \\
    $T_w$ & [s] frequency LP filter time const.                         & 0.01 & 0.01  \\
    $T_r$ & [s] RoCoF washout filter time const.                        & 0.01  & 0.01   \\
    $\Delta t_w$ & [s] time window for RoCoF est.                       & 0.5 & 0.25  \\
    $\epsilon$ & [pu] circulation threshold                                &   -   & 0.05   \\
    $\Delta t_{\Gamma'}$ & [s] Initial time span where $|\Gamma'| > \epsilon$ & 0.0  & 0.023   \\
    
    $d\omega_1$ & [pu/s] $1^{st}$ UFLS threshold                           & 0.012 & 0.012  \\
    $d\omega_2$ & [pu/s] $2^{nd}$ UFLS threshold                           & 0.024 & 0.024   \\
    $\Delta t_{\delta,1}$ & [s]  trip delay for $1^{st}$ UFLS threshold    & 0.2 & 0.2  \\
    $\Delta t_{\delta,2}$ & [s] trip delay for $2^{nd}$ UFLS threshold     & 0.2 & 0.2   \\
    $\Delta LS_1$ & [pu] Load shed by the $1^{st}$ UFLS threshold          & 0.2 & 0.2  \\
    $\Delta LS_2$ & [pu] Load shed by the $2^{nd}$ UFLS threshold          & 0.2  & 0.2   \\
    
    \hline
  \end{tabular}}
  \label{tab:data}
\end{table}

\subsection{\ac{ufls} ROCOF performance results}

\begin{figure}[htb]
    \centering
    \includegraphics[width=0.8\linewidth]{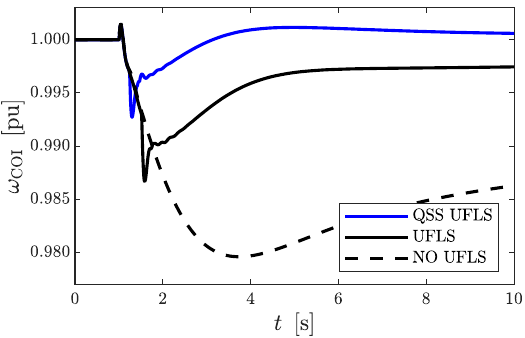}
    \caption{\ac{coi} frequency response for different \ac{ufls} schemes with the specific parameters from Table \ref{tab:data}.}
    \label{fig:wcoi}
    \vspace{-2mm}
\end{figure}

The results of the simulations are depicted in Fig.~\ref{fig:wcoi}.  The three curves represent the frequency of the \ac{coi} for three different simulations:

\begin{enumerate}[(i)]
\item \textbf{NO \ac{ufls}}: the \ac{ufls} is disabled.
\item \textbf{\ac{ufls}}: the \ac{rocof}-based \ac{ufls} scheme is enabled.
\item \textbf{QSS \ac{ufls}}: the QSS \ac{rocof}-based \ac{ufls} scheme is enabled.
\end{enumerate}

\begin{figure}[htb]
    \centering
    \includegraphics[width=0.8\linewidth]{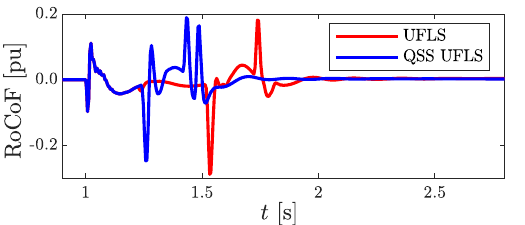}
    \caption{\ac{rocof} estimation results, at bus 20, by means of frequency time derivative for conv.  and QSS UFLS schemes.}
    \label{fig:dw_UFLS_QSS}
    \vspace{-2mm}
\end{figure}

\begin{figure}[htb]
    \centering
    \includegraphics[width=0.8\linewidth]{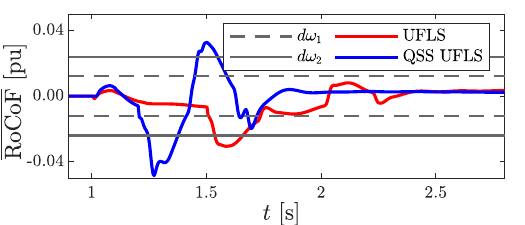}
    \caption{\ac{rocof} estimation results, at bus 20, by means of the average of frequency time derivative for conv.  and QSS UFLS schemes.  Threshold values, $d\omega_1$ and $d\omega_2$, are represented by a black line and a black dotted line respectively.}
    \label{fig:ROCOF_UFLS_QSS}
    \vspace{-2mm}
\end{figure}

\ac{ufls} protection scheme, effectively, improves the system response in terms of frequency stability. The \ac{qss} \ac{rocof}-based \ac{ufls} sheds more load and earlier than the conventional \ac{ufls} logic. This is because on the one hand \acp{rocof} vary among the buses and on the other hand, the conventional \ac{ufls} estimation is slower and at some load buses does not exceed the thresholds due to the larger window.  This behavior can be seen in Figs.  \ref{fig:dw_UFLS_QSS} and \ref{fig:ROCOF_UFLS_QSS} that show, respectively, the time derivative of frequency and average \ac{rocof} estimations at bus 20. This case illustrates that the accuracy and speed of \ac{rocof} estimation could be critical for \ac{ufls} after all if the conventional estimation and logic falls short.

\section{Remarks}\label{sec:remarks}

\subsection{On the Practical Implications}

When estimating the \ac{rocof}, choosing a suitable time window involves a trade-off between measurement accuracy and response promptness.  Although an optimal window length may exist within this balance, \ac{ufls} devices lack knowledge of the nature of the threshold violation event.  For this reason, typically prefixed conventional values are used.

By delaying the \ac{rocof} computation, the optimal time window can be shortened.  However, delaying too much can cause an insufficient or entirely absent response.  Hence, the choice of a time delay is nontrivial and should not be established beforehand, as it depends on the nature of the contingency.  

The proposed omission of \ac{rocof} values that do not fulfill condition ~\eqref{metric} right after the contingency, inherently introduces a delay that affects the \ac{ufls} response.  The introduction of such delay, by means of $\Gamma'$, ensures that the \ac{rocof} estimation is delayed properly; only non-representative values of \ac{rocof} are omitted.  Consequently, averaging time window and tripping delays can be reduced, thereby improving the protection response.  Moreover, as demonstrated in Section \ref{sec:real_data}, the estimation of the \ac{rocof} by means of \ac{qss} frequency is less affected by noise than conventional estimations based on instantaneous frequency, thus enabling faster estimation without compromising accuracy.  

\subsection{On the election of threshold $\epsilon$ }

The need of defining $\epsilon$ comes from potential numerical errors and the discrete nature of the measurements.  If $\epsilon$ is too restrictive (too small), the tout function, $\tout$, might continuously detect noise or estimation imperfections as transient events that lack periodicity, being the edge case equivalent, in the practical application, to not applying the \ac{ufls} protection.  Conversely, if $\epsilon$  is too permissive (too large), $\tout$ might not detect such events and frequency response will be equivalent as conventional \ac{rocof} estimation-based \ac{ufls} protection response.  From this perspective, conventional \ac{ufls} schemes can be seen as a specific case of the \ac{qss}-based scheme with a, \textit{de facto}, over-dimensioned threshold for the system requirements.  

\begin{figure}[htb]
  \centering
  \includegraphics[scale=0.8]{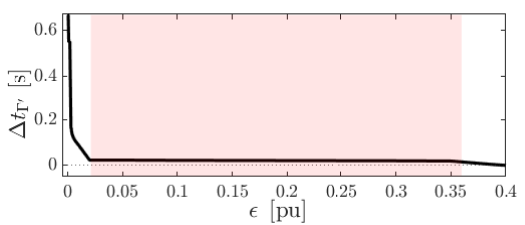}
  \caption{Initial time span where condition \eqref{metric} is not satisfied with respect of different  threshold, $\epsilon$, values.  Range of \ac{qss} \ac{rocof}-based \ac{ufls} optimal performance is red-shadowed.}
  \label{fig: initial_delay_vs_epsilon}
\end{figure}

The choice of $\epsilon$ is guided by experimental observation.  As shown in \cite{gutierrezflorensa2025qss}, transient events that lack periodicity represent an abrupt change of $\Gamma'$, making the discernment of events, with respect to their nature, quite evident.  To support this statement we plot, in Fig~\ref{fig: initial_delay_vs_epsilon}, the initial time delay, $\Delta t_{\Gamma'}$, with respect of different $\epsilon$ values, of the practical application case study.

Within the edge cases, the vast majority of the range of possible $\epsilon$ values falls in a plateau where the delay, $\Delta t_{\Gamma'}$, does not change.
Following the findings in \cite{gutierrezflorensa2025qss}, the $\epsilon$ value, of a specific model, should be chosen only based on the sampling time of the discrete data.  From the experience obtained from real-world data, we recommend to evaluate $\epsilon$ considering measurements taken in stationary conditions and set a quite over-dimensioned value, \textit{e.g.} one order of magnitude higher.  

\section{Conclusions}\label{sec: conclusions}

This paper discusses theoretical and experimental limitations of the \ac{rocof} computation and proposes a novel definition of \ac{rocof}, which is more compliant with the challenges issued by the actual structure of the power system.
Thanks to the definition of the \ac{qss} frequency and the time derivative of the circulation, the time intervals where the frequency has no physical meaning are excluded from the moving average of the classical \ac{rocof} computation.  

This definition shows a significant improvement in the \ac{rocof} estimation when applied to real-world measurement, helping to avoid possible unintentional tripping of \ac{rocof}-based relays during highly-distorted transient events.  
Moreover, if applied to a \ac{rocof}-based \ac{ufls} scheme, robust frequency estimation, by means of $\wqss$, and information from $\Gamma'$ can be used to adapt and shorten the typical rolling window and time delay before the issuing of the relay, speeding up the frequency restoration and decreasing the absolute value of the frequency nadir.  


\begin{thebibliography}{10}
\providecommand{\url}[1]{#1}
\csname url@samestyle\endcsname
\providecommand{\newblock}{\relax}
\providecommand{\bibinfo}[2]{#2}
\providecommand{\BIBentrySTDinterwordspacing}{\spaceskip=0pt\relax}
\providecommand{\BIBentryALTinterwordstretchfactor}{4}
\providecommand{\BIBentryALTinterwordspacing}{\spaceskip=\fontdimen2\font plus
\BIBentryALTinterwordstretchfactor\fontdimen3\font minus
  \fontdimen4\font\relax}
\providecommand{\BIBforeignlanguage}[2]{{%
\expandafter\ifx\csname l@#1\endcsname\relax
\typeout{** WARNING: IEEEtran.bst: No hyphenation pattern has been}%
\typeout{** loaded for the language `#1'. Using the pattern for}%
\typeout{** the default language instead.}%
\else
\language=\csname l@#1\endcsname
\fi
#2}}
\providecommand{\BIBdecl}{\relax}
\BIBdecl

\bibitem{IEEE2018synchrophasors}
``{IEEE/IEC International Standard - Measuring relays and protection equipment
  - Part 118-1: Synchrophasor for power systems - Measurements},'' pp. 1--78,
  2018.

\bibitem{entso2018rocof}
{ENTSO-E}, ``{Rate of change of frequency (RoCoF) withstand capability, ENTSO-E
  guidance document for national implementation for network codes on grid
  connection},'' Brussels, Blegium, Tech. Rep., 2018.

\bibitem{milano2022geometrical}
F.~Milano, ``{A Geometrical Interpretation of Frequency},'' \emph{{IEEE} Trans.
  Power Syst.}, vol.~37, no.~1, pp. 816--819, 2022.

\bibitem{gutierrezflorensa2025qss}
J.~Guti{\'e}rrez-Florensa, {\'A}.~Ortega, L.~Sigrist, and F.~Milano, ``Quasi
  steady-state frequency,'' \emph{{IEEE Trans.~on Circuits and Systems I:
  Regular Papers, \textit{early access}}}, pp. 1--13, 2025.

\bibitem{rocof_meas}
A.~Riepnieks and H.~Kirkham, ``Rate of change of frequency measurement,'' in
  \emph{Int.~Scientific Conference on Power and Electrical Engineering of Riga
  Technical University (RTUCON)}, 2016, pp. 1--5.

\bibitem{rocof_review}
X.~Deng \emph{et~al.}, ``{Review of RoCoF Estimation Techniques for Low-Inertia
  Power Systems},'' \emph{Energies}, vol.~16, no.~9, 2023.

\bibitem{rocof_pll}
R.~Ferrero, P.~A. Pegoraro, and S.~Toscani, ``Proposals and analysis of space
  vector-based phase-locked-loop techniques for synchrophasor, frequency, and
  {ROCOF} measurements,'' \emph{IEEE Trans.~on Instr.~and Meas.}, vol.~69,
  no.~5, pp. 2345--2354, 2020.

\bibitem{phasor_fdt}
M.~Bertocco \emph{et~al.}, ``{Compressive Sensing of a Taylor-Fourier
  Multifrequency Model for Synchrophasor Estimation},'' \emph{IEEE Trans.~on
  Instr.~and Meas.}, vol.~64, no.~12, pp. 3274--3283, 2015.

\bibitem{dft'19}
A.~K. Singh and B.~C. Pal, ``{Rate of Change of Frequency Estimation for Power
  Systems Using Interpolated DFT and Kalman Filter},'' \emph{IEEE Transactions
  on Power Systems}, vol.~34, no.~4, pp. 2509--2517, 2019.

\bibitem{rocof_step}
P.~S. Wright \emph{et~al.}, ``{Field Measurement of Frequency and ROCOF in the
  Presence of Phase Steps},'' \emph{IEEE Trans.~on Instr.~and Meas.}, vol.~68,
  no.~6, pp. 1688--1695, 2019.

\bibitem{rocof_syncrophasor_ULFS}
Y.~Zuo, G.~Frigo, A.~Dervi{\v{s}}kadi{\'c}, and M.~Paolone, ``{Impact of
  Synchrophasor Estimation Algorithms in ROCOF-Based Under-Frequency
  Load-Shedding},'' \emph{IEEE Trans.~on Power Systems}, vol.~35, no.~2, pp.
  1305--1316, 2020.

\bibitem{testrocof}
G.~Rietveld, P.~S. Wright, and A.~J. Roscoe, ``{Reliable
  Rate-of-Change-of-Frequency Measurements: Use Cases and Test Conditions},''
  \emph{IEEE Trans.~on Instr.~and Meas.}, vol.~69, no.~9, pp. 6657--6666, 2020.

\bibitem{pmu_rocof}
G.~Frigo, A.~Dervi{\v{s}}kadi{\'c}, Y.~Zuo, and M.~Paolone, ``{PMU-Based ROCOF
  Measurements: Uncertainty Limits and Metrological Significance in Power
  System Applications},'' \emph{IEEE Trans.~on Instr.~and Meas.}, vol.~68,
  no.~10, pp. 3810--3822, 2019.

\bibitem{rocof_std}
``{IEEE} standard for synchrophasor measurements for power systems -- amendment
  1: Modification of selected performance requirements,'' pp. 1--25, 2014.

\bibitem{entso2016criteria}
{ENTSO-E RG-CE System Protection \& Dynamics Sub Group}, ``{Frequency Stability
  Evaluation Criteria for the Synchronous Zone of Continental Europe},''
  Brussels, Belgium, Tech. Rep., 2016.

\bibitem{rocofislading}
P.~Gupta, R.~S. Bhatia, and D.~K. Jain, ``{Active ROCOF Relay for Islanding
  Detection},'' \emph{{IEEE} Trans. Power Del.}, vol.~32, no.~1, pp. 420--429,
  2017.

\bibitem{rocofLoM}
D.~M. Laverty, R.~J. Best, and D.~J. Morrow, ``Loss-of-mains protection system
  by application of phasor measurement unit technology with experimentally
  assessed threshold settings,'' \emph{IET Generation, Transmission \&
  Distribution}, vol.~9, no.~2, pp. 146--153, 2015.

\bibitem{entso2010technical}
{ENTSO-E System Protection and Dynamics Sub Group}, ``{Technical Background and
  Recommendations for Defence Plans in the Continental Europe Synchronous
  Area},'' Brussels, Belgium, Tech. Rep., 2010.

\bibitem{UFLS_COI}
M.~Sun, G.~Liu, M.~Popov, V.~Terzija, and S.~Azizi, ``{Underfrequency Load
  Shedding Using Locally Estimated RoCoF of the Center of Inertia},''
  \emph{{IEEE} Trans. Power Syst.}, vol.~36, no.~5, pp. 4212--4222, 2021.

\bibitem{kundur2007power}
P.~Kundur \emph{et~al.}, ``Power system stability,'' \emph{Power system
  stability and control}, vol.~10, no.~1, pp. 7--1, 2007.

\bibitem{spain}
\BIBentryALTinterwordspacing
{ENTSO-E}. (2025) \BIBforeignlanguage{en-us}{28 {April} {Blackout}}. [Online].
  Available:
  \url{https://www.entsoe.eu/publications/blackout/28-april-2025-iberian-blackout/}
\BIBentrySTDinterwordspacing

\bibitem{chile}
\BIBentryALTinterwordspacing
G.~Lemos, A.~Melgar, M.~Torres, and M.~Rios. \BIBforeignlanguage{en}{State of
  emergency declared after blackout plunges most of {Chile} into darkness}.
  [Online]. Available:
  \url{{edition.cnn.com/2025/02/25/americas/chile-blackout-14-regions-intl-latam}}
\BIBentrySTDinterwordspacing

\bibitem{ulfs_scheme}
U.~Rudez and R.~Mihalic, ``Wams-based underfrequency load shedding with
  short-term frequency prediction,'' \emph{{IEEE} Trans. Power Del.}, vol.~31,
  no.~4, pp. 1912--1920, 2016.

\bibitem{entso2020rocof}
{ENTSO-E}, ``{Inertia and Rate of Change of Frequency (RoCoF)},'' ENTSO-E,
  Brussels, Belgium, Tech. Rep. v17, 2020.

\bibitem{milano2025lagrange}
F.~Milano, ``{Equivalence Between Geometric Frequency and Lagrange
  Derivative},'' \emph{{IEEE} Trans. Circuits Syst. {I}}, vol.~72, no.~9, pp.
  4800--4809, 2025.

\bibitem{milano2022paradoxes}
F.~Milano \emph{et~al.}, ``Using differential geometry to revisit the paradoxes
  of the instantaneous frequency,'' \emph{IEEE Open Access Journal of Power and
  Energy}, vol.~9, pp. 501--513, 2022.

\bibitem{truesdell1954kinematics}
C.~Truesdell, \emph{The Kinematics of Vorticity}.\hskip 1em plus 0.5em minus
  0.4em\relax Indiana, US: Indiana University Press, 1954.

\bibitem{Stoker1969differential}
J.~J. Stoker, \emph{Differential Geometry}.\hskip 1em plus 0.5em minus
  0.4em\relax John Wiley \& Sons, Ltd, 1969.

\bibitem{rocof500}
A.~Dervi{\v{s}}kadi{\'c}, Y.~Zuo, G.~Frigo, and M.~Paolone, ``{Under Frequency
  Load Shedding based on PMU Estimates of Frequency and ROCOF},'' in \emph{IEEE
  ISGT-Europe}, 2018, pp. 1--6.

\bibitem{DER}
G.~Tzounas and F.~Milano, ``Improving the frequency response of {DERs} through
  voltage feedback,'' in \emph{IEEE PES General Meeting}, 2021, pp. 1--5.

\end{thebibliography}



\end{document}